\documentclass[acmlarge,screen]{acmart}

\usepackage{booktabs} % For formal tables

\usepackage[ruled]{algorithm2e} % For algorithms
\usepackage{paralist}
\usepackage{listings}
\usepackage{pgfplotstable}
\usepackage{pgfplots}
\usepackage{hyperref}

\SetAlFnt{\small}
\SetAlCapFnt{\small}
\SetAlCapNameFnt{\small}
\SetAlCapHSkip{0pt}
\IncMargin{-\parindent}

\usepackage{xpatch}

\makeatletter
\xpatchcmd{\ps@firstpagestyle}{Manuscript submitted to ACM}{}{\typeout{First patch succeeded}}{\typeout{first patch failed}}
\xpatchcmd{\ps@standardpagestyle}{Manuscript submitted to ACM}{}{\typeout{Second patch succeeded}}{\typeout{Second patch failed}}    \@ACM@manuscriptfalse% Also in titlepage
\makeatother

\renewcommand\footnotetextcopyrightpermission[1]{} % removes footnote with conference info

\setcopyright{rightsretained}
\usepackage{color}

\definecolor{TUMblue}{RGB}   {  0, 101, 189} % "Pantone300"
\definecolor{Pantone540}{RGB}{  0,  51,  89}
\definecolor{Pantone301}{RGB}{  0,  82, 147}
\definecolor{Pantone285}{RGB}{  0, 115, 207}
\definecolor{Pantone542}{RGB}{100, 160, 200}
\definecolor{Pantone283}{RGB}{152, 198, 234}
\definecolor{TUMdarkgray}{RGB}{ 88, 88, 90}
\definecolor{TUMgray}{RGB}{ 156, 157, 159}
\definecolor{TUMlightgray}{RGB}{ 217, 218, 219}
\definecolor{TUMgreen}{RGB}{162, 173, 0} % Pantone383
\definecolor{TUMorange}{RGB}{227, 114, 34} % Pantone158
\definecolor{TUMelfenbein}{RGB}{218, 215, 203} %Pantone7527
\definecolor{TUMpyellow}{RGB}{255, 180, 0}
\definecolor{TUMporange}{RGB}{255, 128, 0}
\definecolor{TUMpred}{RGB}{229, 52, 24}
\definecolor{TUMpdarkred}{RGB}{202, 33, 63}
\definecolor{TUMpblue}{RGB}{0, 153, 255}
\definecolor{TUMplightblue}{RGB}{65, 190, 255}
\definecolor{TUMpgreen}{RGB}{145, 172, 107}
\definecolor{TUMplightgreen}{RGB}{181, 202, 130}
\definecolor{TUMpbluethemelower}{RGB}{0, 82, 147} %=Pantone301
\definecolor{TUMpbluethemeupper}{RGB}{0, 40, 72}

\definecolor{shadecolor}{gray}{.95}                                                           %farbe für listings

\begin{document}
% Title portion. Note the short title for running heads
\title[valgrind-ws]{A Valgrind Tool to Compute\\the Working Set of a Software Process}

\author{Martin Becker}
\orcid{0000-0003-3195-0503}
%\email{martin.becker@tum.de}
\author{Samarjit Chakraborty}
\affiliation{%
  \institution{Chair of Real-Time Computer Systems\\Technical University of Munich}
  \streetaddress{Arcisstrasse 21}
  \city{Munich}
%  \state{VA}
  \postcode{80333}
  \country{Germany}}
%\email{samarjit@tum.de}

\begin{abstract}
  This paper introduces a new open-source tool for the dynamic
  analyzer \emph{Valgrind}.  The tool measures the amount of memory
  that is \emph{actively} being used by a process at any given point
  in time.  While there exist numerous tools to measure the memory
  requirements of a process, the vast majority only focuses on metrics
  like resident or proportional set sizes, which include memory that
  was once claimed, but is momentarily disused. Consequently, such
  tools do not permit drawing conclusions about how much cache or RAM
  a process actually requires at each point in time, and thus cannot
  be used for performance debugging.  The few tools which do measure
  only actively used memory, however, have limitations in temporal
  resolution and introspection.  In contrast, our tool offers an easy
  way to compute the memory that has recently been accessed at any
  point in time, reflecting how cache and RAM requirements change over
  time.  In particular, this tool computes the set of memory
  references made within a fixed time interval before any point in
  time, known as the \emph{working set}, and captures call stacks for
  interesting peaks in the working set size.  We first introduce the
  tool, then we run some examples comparing the output from our tool with
  similar memory tools, and we close with a discussion of limitations.
\end{abstract}




\keywords{Process, Memory Consumption, Working Set, Simulation}

\begin{teaserfigure}
  \centering
  \includegraphics[width=.9\textwidth]{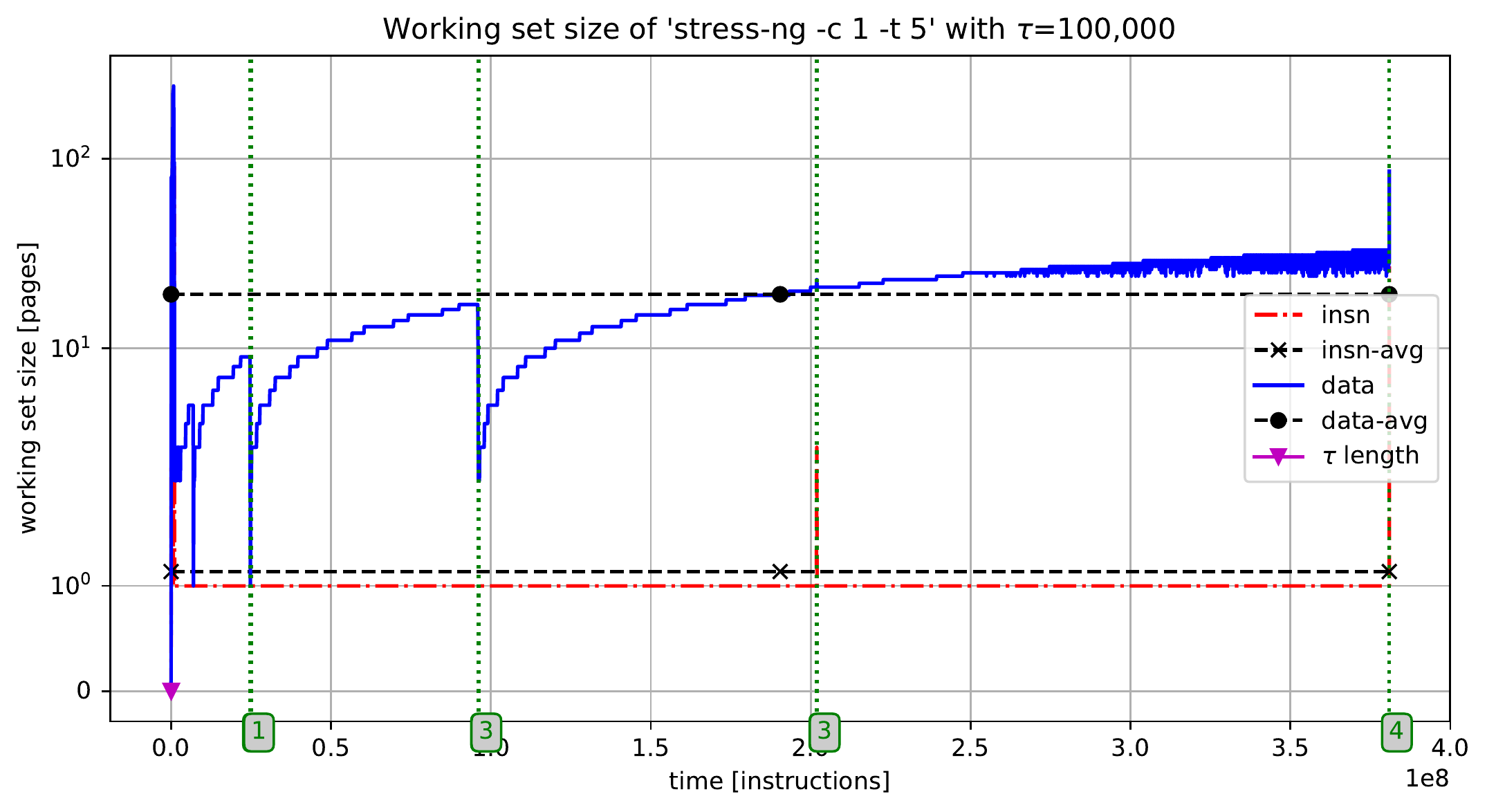}
  \caption{Visualization of the working set, generated from tool output.}
  \label{fig:teaser}
\end{teaserfigure}

\maketitle
\newpage
\section{Introduction}
The \emph{Memory Wall}~\cite{DBLP:journals/sigarch/WulfM95} is
arguably one of the biggest performance challenges in modern computer
systems. Since the speed gap between processors and memory is
currently several orders of magnitude and still diverging, it
becomes more and more important to understand the memory bottlenecks
of an application.

A na\"ive approach to measuring memory requirements would be to
determine the total amount of memory that an application has
claimed. While this can be useful as a first impression, it does not
give any information about how much of that memory is actively being
used. The principle of (temporal) locality tells us, that recently
used memory is also more likely to be used soon in the
future. Therefore, although an application may have occupied a large
amount of memory, it is likely to only operate on a subset at any
point in time. This, in turn, is important for the memory hierarchy:
Fast memory needs to be small, and thus lower memory requirements will
make a faster application. In summary, we should not be concerned with
the total amount of memory that an application has claimed, but only
whether we can bring the soon-to-be-required memory contents quickly
enough into the fast memory.

In 1968, Peter Denning~\cite{denning1968working} introduced the
concept of a \emph{working set}, as a means to unify computation and
memory management. It represents a process' demand for memory, tightly
coupled to the computation being performed. Specifically, at any point
in time $t$, it is defined as the memory that has been accessed within
the last $\tau$ time units, also known as the \emph{working set
  window}. As such, the working set does not include momentarily
disused memory, and thus its size is usually much smaller than the
total number of pages mapped (``resident set size''). It also directly
translates into how much precious cache and RAM is required in that
time interval, and thus also allows for estimating memory throughput
requirements to keep a program running. We refer the reader to
Denning's paper for a formal definition of the working set.

Surprisingly, while there are countless tools to measure resident set
size (e.g., Linux' \texttt{top}, Windows' task manager) or virtual
memory (same tools as above), only few tools are available today to measure the
working set. One possible explanation is that estimating the working
set would require tracking memory accesses, which can be costly and
have side-effects. In fact, as long as no page faults occur, an
operating system would not even be capable of seeing page
accesses. Consequently, such tools often invalidate MMU entries and
observe how they reappear over time, and thus they are intrusive.  One
recent, non-intrusive tool to measure the working set under Linux is
\texttt{wss}~\cite{wss18}. It builds on kernel patches that were added
in 2015~\cite{lkml15}, which allow tracking of idle pages, without
changing the state of the memory subsystem. This tool only works for
Linux newer than 4.3, and is not enabled in mainline
builds. Furthermore, it has several minor limitations, such as only
tracking pages on the LRU list and only working with certain
allocators (the \emph{slab} allocator does not use an LRU list).  Most
importantly, in contrast to our tool, it does not give any introspection
beyond the size of the working set, due to restrictive performance
penalties for online implementations. %Further tools are summarized in
%Section~\ref{sec:related-work}.
% FIXME: are these tools tracking instruction memory, as well?

To offer a more generic and precise way of measuring the working set
for both instruction and data memory, we have developed a tool for the
popular instrumentation framework
Valgrind~\cite{nethercote2007valgrind}. It seamlessly integrates into
the latest Valgrind version, and thus works on many Linux platforms,
such as x86, ARM/Android, MIPS32, etc. Computing the working set is
achieved by tracking all memory accesses of a process (both
instruction and data), and counting the number of pages that have been
accessed within the last $\tau$ (user-defined) time units at
equidistant sampling points $t=kT$, with $k\in\mathbb N$ and $T$ being
the sampling interval. The output is the size of the instruction and
data working sets over time, for each individual thread, annotated
with additional information.

\pagebreak
\section{Tool Details}
Our tool for measuring the working set is made open source, and
available at 
%\url{https://revealed-after-acceptance.org}.
\url{https://github.com/mbeckersys/valgrind-ws}. 
It tracks the number of page accesses over time, calculates the
working set size (WSS) individually for each process, and separately
for data and instruction memory. It is a plugin for
Valgrind~\cite{nethercote2007valgrind}, and thus the program under
analysis is effectively being instrumented and simulated, while using
the host ISA.

\subsection{Interaction with Valgrind Core}
\begin{figure}[htb]
  \centering
  \includegraphics[width=.5\textwidth]{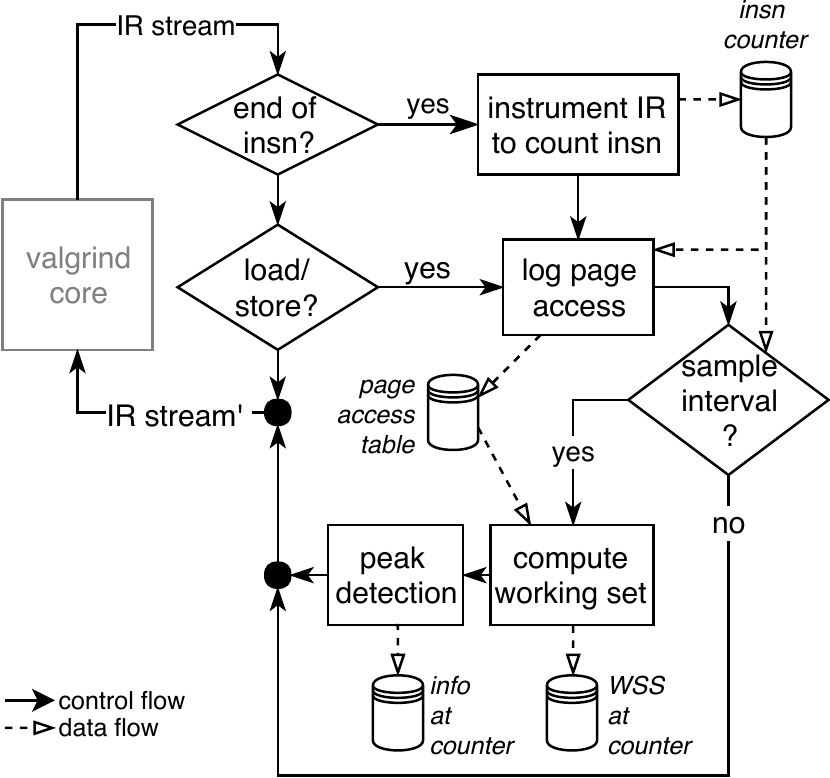}
  \caption{Interaction with Valgrind core\label{fig:internals}}
  \vspace*{-1em}
\end{figure}

Valgrind tools work by instrumenting and observing a process that is
interpreted by the Valgrind core. In particular, the Valgrind core
provides a stream of an architecture-independent intermediate
representation (IR), which carries markers relating the IR to the
original instruction stream. Figure~\ref{fig:internals} shows how we
interact with the core. We log all page references for both
instruction and data, and count recently accessed pages, i.e., the
WSS, at equidistant time intervals.

The time intervals are based on the number of instructions simulated,
because the simulation slows down the process under analysis, and thus
wall-clock time would be meaningless. Towards that, we instrument the
IR stream with IR statements that increment a counter every time an
instruction marker is reached. Page accesses are registered only by
observing the incoming IR statements. For every data access and every
instruction marker (indicating a new original instruction) we query
the referenced addresses, translate them to page addresses, and
maintain a table holding all seen pages together with a time stamp of
the last access.

Every time the instruction counter reaches a multiple of the sampling
interval $T$, we compute the working set by counting all pages that
have been accessed within the last $\tau$ instructions. Additionally, a
peak detection algorithm can be enabled, which records additional
information when the WSS exhibits peaks, and is described in the
following section.

\subsection{Online Peak Detection and Annotation}
For a developer it is important to understand why her program exhibits
a certain WSS behavior so that it can be analyzed and improved.
Therefore, additional information about the WSS samples, such as the
current call stack, shall be recorded.  However, this cannot be done
for every sample, since it would significantly increase the memory
footprint of our tool, and thus slow down the simulation
unfavorably. Thus, the tool detects peaks in the WSS of either
instructions or data, and records additional information only for
those peaks. Such additional information is indicated in the output by
boxed numbers, e.g., \fbox{1} and \fbox{3}, see
Figure~\ref{fig:teaser}.

Peak detection must be quick to respond to peaks, otherwise we blame
the wrong code location. It also must have a low memory footprint, since it
otherwise fails its own purpose. Therefore, we build on the
\emph{principle of dispersion} and use signal filters that do not
require storing a window of samples. The threshold for peak detection
depends on the current signal properties; if the memory usage is high,
a certain distance from the moving average is used as threshold,
otherwise, we compare the deviation from mean against the variance. In
between, the threshold is a mixture of both.

This strategy allows us
to identify meaningful peaks in both stationary and non-stationary
memory usage settings. Specifically, we first compute an exponential
moving average $\mu$ and moving variance $\sigma^2$; both of them do
not require storing a list of recently seen samples. Then, for every
new sample $x$, we calculate its distance $e = |x - \mu|$, and the
ratio $F=\sigma^2/\mu$, also known as the \emph{Fano factor}. The new
sample $x$ is considered a peak if $e > E$, where $E$ is the
time-varying detection threshold, calculated as $E=cg\sigma^2 +
(1-c)g\mu$, with $c=1-\exp(F/2)$, and $g$ being a parameter to
influence detection sensitivity.

Finally, we apply an exponential filtering to the signal before
updating the moving average/variance, which prevents that single huge
peaks skyrocket these statistics and prevent detection of subsequent
peaks. To satisfy the requirement of quick response to peaks, we only
apply filtering as long as a peak is present.

\subsection{Hot Pages}
The tool also yields a list of the most frequently accessed pages,
both for instruction and data. We use debug information to provide
additional information about the pages, such as source locations.
For example:
\begin{lstlisting}
Code pages (57 entries):
count     page    information
90312000  0x0400  touch_pages (pageramp.c:38)
  112640  0x4D08  __vsyslog_chk (syslog.c:298)
...
\end{lstlisting}
It can be seen that one single page of instructions is the target of
most page accesses. Since we only track at page granularity, the
information here is only approximate. It refers to the first
instruction/data access that falls into the given page. A similar
output is produced for data pages.

%\subsection{Visualization}
%The tool is shipped with a Python script that produces the output
%as shown in Figure~\ref{fig:teaser}.

\section{Examples}
In this section we demonstrate several examples, to illustrate the
tool's output and use cases. We compare our output to the following
tools:
\begin{inparaenum}[(1)]
%\item GNU \texttt{time} (max. RSS, minor page faults),
\item \texttt{psutil}~\cite{psutil} (a Python script to programmatically query process details) and
\item \texttt{wss}~\cite{wss18} (a recent Linux tool to measure the working set size).
\end{inparaenum}
One notable difference is that our simulation has a different time
base than the other two tools. Both of them measure at wall-clock
time, but incur different overhead. Our tool, in contrast, uses the
number of executed instructions as a time basis. Therefore, in a
practical setting, all three tools have different time bases and the
output should not be compared directly. Another difference is that the
other tools are quite limited in their temporal resolution. For the
examples given here, we had them sample the memory state every 2ms,
which already keeps our machine busy for these simple example
programs.

\subsection{``Pageramp'' Demo}
\begin{lstlisting}
valgrind --tool=ws --ws-peak-detect=yes ./pageramp
\end{lstlisting}
This is a simple workload requesting and releasing data pages in a
sawtooth pattern.  The program starts with zero data
pages, %(additionally to whatever non-user data pages it
%may already have), 
and successively claims more and more pages, until an upper bound of
1024 pages is reached. Subsequently, it releases the pages again,
until we arrive back to zero. During the entire time, it writes one
byte of data to every second page, to ensure that half of the claimed
memory is actively used.  This process is repeated 10 times before the
program exits.

\begin{figure}[tb]
  \centering
  \includegraphics[width=.7\textwidth]{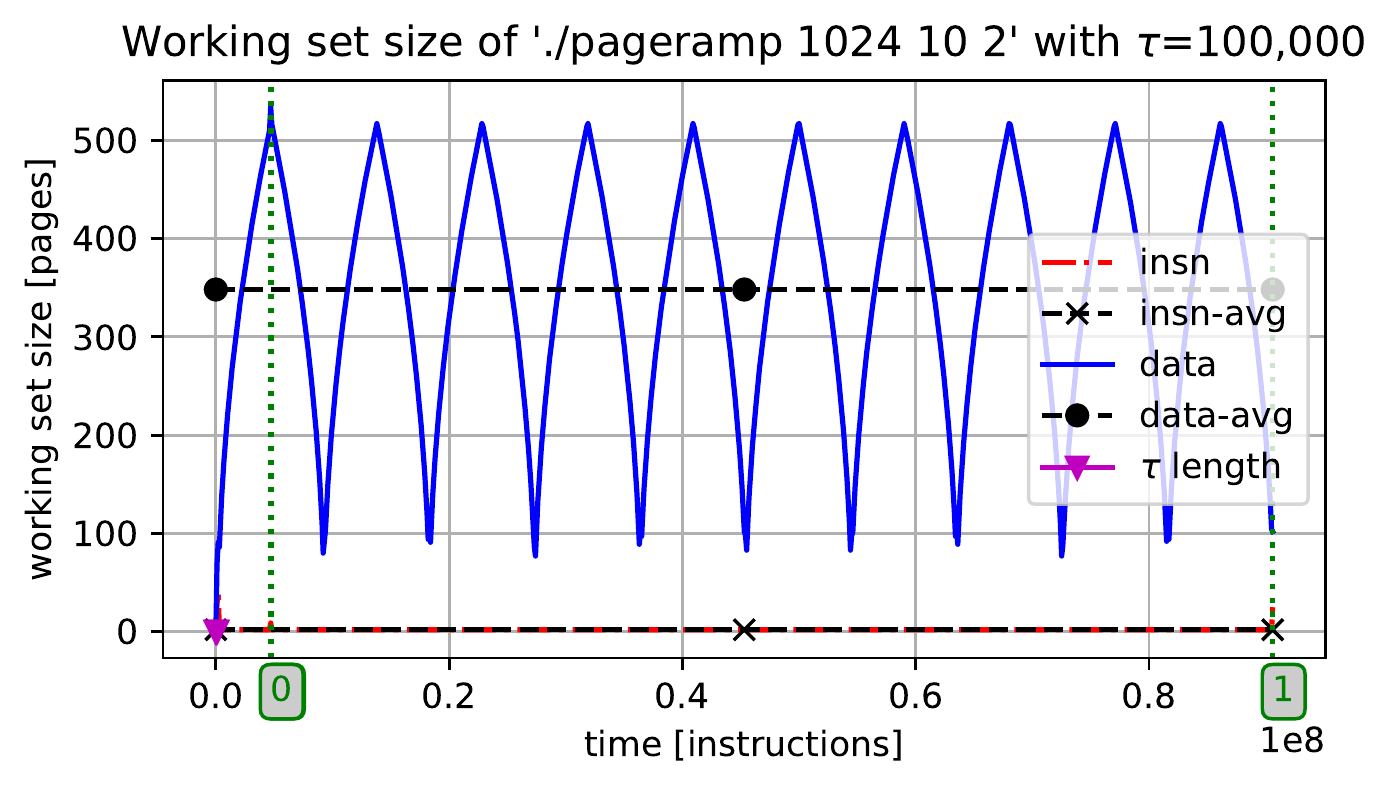}
  \caption{Memory requirements of "pageramp", as reported by our tool ($\tau=100,000$ instructions)\label{fig:pageramp:us}}
\end{figure}

\begin{figure}[tb]
  \pgfplotstableread{examples/pageramp_mod/wss.csv}\wssdata
\pgfplotstableread{examples/pageramp_mod/ps.csv}\psdata
\begin{tikzpicture}[font=\small]
\begin{axis}[
  xlabel={time [s]},
  y label style={at={(0.07,0.5)}},
  ylabel=pages,
  legend style={fill=white, fill opacity=.8},
  grid,
  enlargelimits=false,
  width=1.0*\columnwidth,
  height=5cm,
  x tick label style={
    /pgf/number format/.cd,
    fixed,
    fixed zerofill,
    precision=2,
    /tikz/.cd
  },
]

% wss
%\addplot [color=black, mark=triangle] table [x=Est, y expr=\thisrow{RSS}]{\wssdata};
%\addlegendentry{\texttt{wss}/RSS}
\addplot [color=black, mark=x] table [x=Est, y expr=\thisrow{Ref}]{\wssdata};
\addlegendentry{\texttt{wss}/WSS}
%\addplot table [x=Est, y expr=\thisrow{PSS}]{\wssdata};
%\addlegendentry{\texttt{wss}/PSS}

% psutil
\addplot [color=red, mark=triangle*] table [x=t, y expr=\thisrow{RSS}]{\psdata};
\addlegendentry{\texttt{psutil}/RSS}
%\addplot table [x=t, y expr=\thisrow{vms}]{\psdata};
%\addlegendentry{\texttt{psutil}/VMS}
%\addplot [color=red, mark=+] table [x=t, y expr=\thisrow{data}]{\psdata};
%\addlegendentry{\texttt{psutil}/data}

\end{axis}
\end{tikzpicture}
  \caption{Memory requirements of ``pageramp'' as reported by \texttt{wss} and \texttt{psutil}.}
  \label{fig:pageramp:others}
\end{figure}
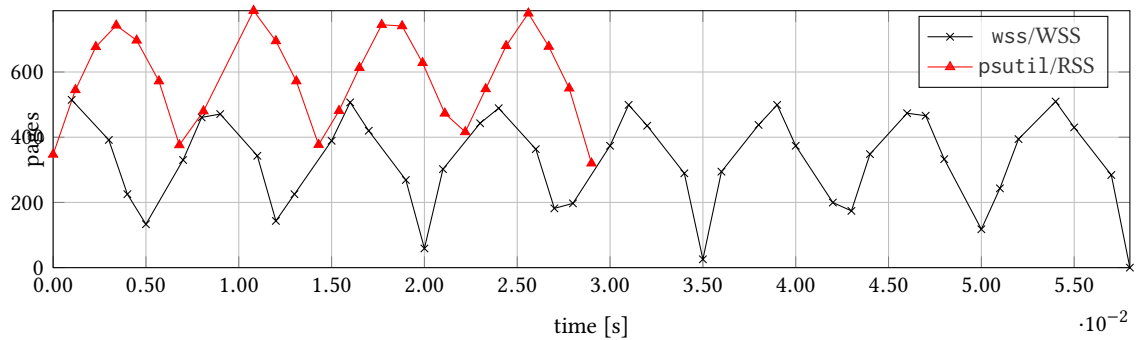

We expect that both our tool and \texttt{wss} tools show similar
output, and that \texttt{psutil} overestimates the memory requirements
roughly by a factor of two.  As it can be seen in
Fig.~\ref{fig:pageramp:us}, our tool clearly yields the expected
output. The output from \texttt{psutil} and \texttt{wss} are depicted
in Fig.~\ref{fig:pageramp:others}. Both programs miss some part of the
execution, since they can only be started after the process has come
to life.  Additionally, they have a worse temporal resolution. More
importantly, while \texttt{wss} delivers the expected result (yet
without any introspection), \texttt{psutil} can only provide an upper
bound on the working set size by means of the resident set size (RSS).
Last but not least, the other tools do not separate between
instruction and data.

Only our tool provides further information about the WSS. Two peaks
have been detected, marked as \fbox{0} and \fbox{1} in the output.
The latter one is a peak in the instruction WSS, due to OS
cleanup actions. The first one is caused by user code, and the information
recorded for \fbox{0} is
\begin{lstlisting}
[0] refs=2, loc=pageramp.c:37|pageramp.c:77
\end{lstlisting}
saying that the call stack at this peak was \texttt{pageramp.c} line
37, called by line 77. The corresponding code is
\begin{lstlisting}
36:  static int touch_pages(pagetab_t *pt, int stride) {
37:     for (int p=0; p<=(pt->num-stride); p+=stride) {
...
77:  touch_pages(&pt, stride);
\end{lstlisting}
and thus the tool is pointing the user directly to the piece of code
that caused the latest access.

Finally, the tool also provides some summaries, which give a first
impression about average (avg), maximum WSS (peak) and the total
number of unique pages being accessed:
\begin{lstlisting}
Insn avg/peak/total: 2.1/40/57 pages (8/160/228 kB)
Data avg/peak/total: 348.5/534/1098 pages (1394/2136/4392 kB)
\end{lstlisting}
Note that our example program apparently uses 74 data pages that are
not requested by the user, but by the C library.

\subsection{Working Set Window $\tau \neq$ Sampling Period $T$}
\begin{lstlisting}
valgrind --tool=ws --ws-tau=10000 ./pageramp
...
valgrind --tool=ws --ws-tau=10000000 ./pageramp
\end{lstlisting}
So far, the working set window $\tau$ has been the same length as the
sampling interval $T$, since the other tools cannot measure when $\tau
\neq T$: \texttt{psutil} does not measure the working set,
so $\tau$ is meaningless here, and \texttt{wss} modifies bits in the
page table every time a sample is taken, and therefore cannot handle
this case. However, it is important for scalability of the
measurements to separate these two aspects. While for long-running
programs the sampling interval should be increased to reduce the
amount of measurement overhead, the working set window should be
chosen according to a maximum memory bandwidth, which is independent
of the program under analysis.

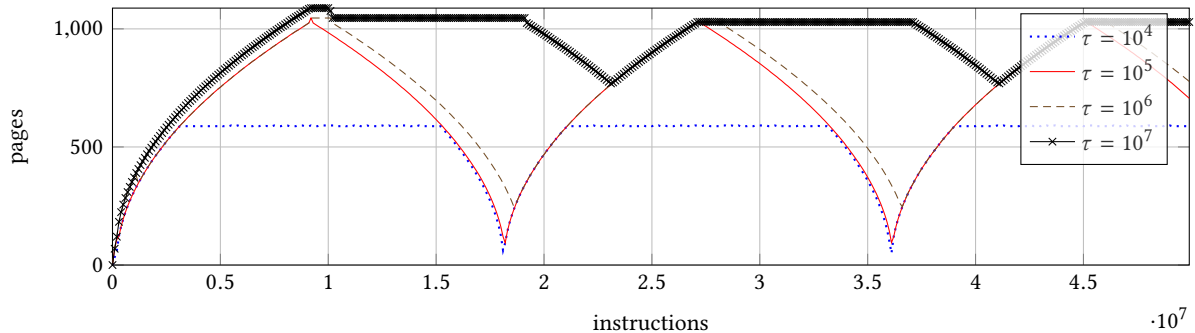
\begin{figure}[tb]
  \centering
  \pgfplotstableread{examples/pageramp/v-ws.10k.csv}\vwstenk
\pgfplotstableread{examples/pageramp/v-ws.100k.csv}\vwshunk
\pgfplotstableread{examples/pageramp/v-ws.1M.csv}\vwsoneM
\pgfplotstableread{examples/pageramp/v-ws.10M.csv}\vwstenM
\begin{tikzpicture}[font=\small]
\begin{axis}[
  restrict x to domain=0:5e7,
  xlabel={instructions},
  ylabel=pages,
  legend style={fill=white, fill opacity=.8},
  grid,
  enlargelimits=false,
  width=1.0*\columnwidth,
  height=5cm,
]

% our tool
\addplot+ [mark=none, thick, dotted] table [x=t, y=WSS_data]{\vwstenk};
\addlegendentry{$\tau=10^4$}
\addplot+ [mark=none] table [x=t, y=WSS_data]{\vwshunk};
\addlegendentry{$\tau=10^5$}
\addplot+ [mark=none, densely dashed] table [x=t, y=WSS_data]{\vwsoneM};
\addlegendentry{$\tau=10^6$}
\addplot+ [mark=none, mark=x, samples=100000] table [x=t, y=WSS_data]{\vwstenM};
\addlegendentry{$\tau=10^7$}
\end{axis}
\end{tikzpicture}\vspace*{-1em}
  \caption{Data working set size of ``pageramp'' as reported by tool for
    different working set windows $\tau$}
  \label{fig:pageramp:tau}\vspace*{-1em}
\end{figure}

We now exercise the test program ``pageramp'' again with our tool,
while choosing different working set windows $\tau$. The results are
shown in Figure~\ref{fig:pageramp:tau}. For $\tau=10,000$
instructions, the working set size never exceeds 600 pages. The pages
cannot be accessed fast enough to include all the 1024 pages in the
working set; or conversely, it takes more than 10,000 instructions to
touch all 1024 pages. For larger $\tau$, the upper peaks become
clearly visible, since now $\tau$ is large enough so that at least
briefly all pages are considered active before they leave the working
set again. In turn, the lower bound of the working set is now
increased, because the already released pages are still considered to
be active. The latter behavior could in principle be fixed by tracking
the release of pages, but this would deviate the definition of the
working set.

\section{Discussion}
\subsection{Performance}
As any Valgrind tool, the execution is slowed down. The slowdown
depends on the number of memory accesses, and also on the sampling
interval. A larger sampling interval should be chosen for long-running
programs, which reduces the memory footprint and the simulation time.

In line with other Valgrind tools, all the results (e.g., the working
set size over time) need to be stored in memory before they can be
written to file. Therefore, our tool's memory requirements increase
with the run-time of the workload. Because the workload itself is
simulated in the Valgrind core, this has no influence on the
output. %If the memory of the host is exhausted, we stop taking samples
%and let the workload finish with a warning.

%Peak detection degrades performance by TODO.

\subsection{Precision}
The sampling interval is not exactly equidistant, but happens only at
the end of superblocks or exit IR statements. Thereby, a small jitter
must be expected, which is usually in the order of a few dozen of
instructions. In our implementation, the jitter is strictly positive,
i.e., the sampling interval given by the user is never undercut.

\subsection{Limitations}
% Pages that are being released (e.g., using \texttt{munmap}), are not
% actively being removed from the working set. While this conforms with
% the definition of the working set, the amount of memory required to
% store the set is slightly overestimated shortly after page releases,
% since immediately after the release the memory could be used
% otherwise.

Sharing pages between threads or processes is currently ignored,
therefore the working set is overestimated for multi-threaded programs
sharing data.

If pages are unmapped and a new page is later mapped under the same
address, they are counted as the same page, even if the contents are
different. This is less critical for the working set size at any
given point in time, but affects the summaries given in the end.

Finally, we want to explicitly warn that OS- and hardware mechanisms,
such as read-ahead and prefetching, are not being considered by virtue
of Valgrind's execution model. This program only counts demand
accesses, as logically required by the running process.

\section{Related Work\label{sec:related-work}}
Most attempts to measure the working set size in a running system are
based on MMU invalidation, as the \texttt{wss} tool~\cite{wss18}
mentioned earlier.  Under the Windows\texttrademark~operating system,
the equivalent tools are the \emph{Windows Performance Recorder} and
\emph{Xperf}~\cite{winrefset}. Such measurements are intrusive, since
the tools actively remove pages from the MMU, and observe how they are
being loaded back again. Also, they cannot track instruction memory.
Another mentionable example for tracking the working set size via MMU
invalidation is
\cite{zhao2011low}. %The authors have found that the working set is
%often stable over larger time intervals, and exploit this to reduce
%tracking overhead. % =dynamic hot set sizing

Gregg~\cite{GreggWss18} mentions other ways to estimate the WSS, among
them the hardware performance monitoring counters (PMCs) in modern
CPUs. Deducing the working set size from the PMCs would require
measuring cache and DRAM in/out traffic, and possibly MMU accesses,
and deducing from that which pages are being used. Since counters are
processor-specific, such an approach would have to be developed for
each processor family individually, and likely not yield precise
results due to lacking implementation details.  Other approaches he
mentions are memory breakpoints (slow), processor traces and bus
snooping (needs hardware).

Another approach for taking WSS measurements is via
virtualization. Since full virtualization or tracking every memory
access would be too slow, these approaches are typically also based on
MMU invalidation. Efficient implementations for such environments have
been presented recently~\cite{wires2014characterizing}.  An approach
based on soft PMCs of the Windows guest operating in a virtualization
environment has recently been proposed
in~\cite{melekhova2015estimating}. However, virtualization in general
requires more setup work than our tool, and thus is less convenient
for daily use.

% An indirect way to measure the working set are \emph{miss ratio
% curves} (MRC). The idea is to observe the caching behavior under
% different cache sizes, and deduce the working set size from that. A
%dynamic MRC measurement has been proposed in ~\cite{zhou2004dynamic}
%% count: 233
%A more efficient implementation would be based on \emph{counter
%  stacks}, 

%Alternative algorithms for peak detection are evaluated by
%Palshikar~\cite{palshikar2009simple}. % no info in it?

The tool presented here falls under the simulation category, and thus
offers more insights than the mentioned methods, for the price of
slower execution. It is, however, a generic tool for an
instrumentation framework that is widely used on many Linux platforms,
and does not depend on the caching hardware. We are not aware of any
publicly available tool with similar functionality.

\section{Closing Remarks}
We have introduced a new tool for the popular Valgrind suite, which
determines the active memory requirements of a process, known as the
\emph{working set size}, and allows associating peaks in the working
set size with call stack information. The measurements are taken at
page-level granularity, at a user-defined interval and working set
window. This tool can be used to monitor the time-varying memory
requirements of an application, and subsequently leverage this
information for performance debugging.

This first release has some limitations that need to be considered by
a user. At the moment, we do not track release and re-use of pages,
and we do not consolidate the working sets of several threads in
multi-threaded applications. All of that results in a slight
overapproximation of the working set, especially when shared memory is
used. Future work entails addressing these shortcomings, and possibly
also introducing models to measure spatial locality of memory
accesses.

\bibliographystyle{ACM-Reference-Format}
\bibliography{paper}

\end{document}